# A Composable Just-In-Time Programming Framework with LLMs and FBP

Andy Vidan
Composable Analytics, Inc.
Cambridge, MA USA
andy@composableanalytics.com

Lars Fiedler
Composable Analytics, Inc.
Cambridge, MA USA
lars@composableanalytics.com

*Abstract*— **This paper introduces a computing framework that combines Flow-Based Programming (FBP) and Large Language Models (LLMs) to enable Just-In-Time Programming (JITP). JITP empowers users, regardless of their programming expertise, to actively participate in the development and automation process by leveraging their task-time algorithmic insights. By seamlessly integrating LLMs into the FBP workflow, the framework allows users to request and generate code in real-time, enabling dynamic code execution within a flow-based program. The paper explores the motivations, principles, and benefits of JITP, showcasing its potential in automating tasks, orchestrating data workflows, and accelerating software development. Through a fully implemented JITP framework using the Composable platform, we explore several examples and use cases to illustrate the benefits of the framework in data engineering, data science and software development. The results demonstrate how the fusion of FBP and LLMs creates a powerful and user-centric computing paradigm.**

## I. INTRODUCTION

In the ever-evolving landscape of software development, traditional approaches often struggle to keep up with dynamic user requirements. To address this challenge, we introduce the concept of Just-In-Time Programming (JITP) and develop a JITP computing framework by integrating Flow-Based Programming (FBP) and Large Language Models (LLMs). This paper explores the motivations, principles, and benefits of JITP, showcasing its potential for orchestrating complex data workflows, building task automation processes and developing software solutions. We demonstrate a fully implemented JITP framework using the Composable DataOps Platform [1] and show that flow-based programming techniques and large language models provide a powerful framework for end-users to implement algorithms during task execution.

To motivate and provide context for Just-In-Time Programming, we first consider Just-In-Time Analytics. Just-In-Time Analytics is an approach that focuses on providing end-users with the ability to query and interrogate data directly to build analytical products that contain timely and relevant insights. It involves the real-time analysis of data and the delivery of actionable information precisely when it is needed during their decision-making processes. By minimizing delays and optimizing data processing and delivery, Just-In-Time Analytics enables users to make informed decisions based on up-to-date and contextually appropriate information.

Just-In-Time Analytics has demonstrated the value of real-time insights in data analysis, empowering organizations to make timely and informed decisions. This success highlights the potential for a similar concept in programming. Traditional programming methods often require extensive upfront planning and design, which may not align with the dynamic nature of user tasks. Just-In-Time Programming aims to bridge this gap by allowing users to implement algorithms during task execution, aligning software functionality with immediate needs.

Just-In-Time Programming is centered around empowering users to actively engage in programming and task automation during task execution, rather than relying solely on pre-designed software solutions [2]. In this paradigm, the user takes on a central role, irrespective of their programming expertise or experience level. Whether a novice user or an experienced programmer, individuals can benefit from a JITP approach.

For novice users, JITP offers a user-friendly and accessible entry point into programming. It enables them to recognize algorithmic opportunities and implement computing solutions in real-time, without the need for extensive prior programming knowledge. By embracing JITP, novices can leverage their domain expertise and insights gained during task execution to create software solutions tailored to their specific needs without being constrained by the limitations of pre-designed software.

Experienced programmers can also find value in JITP as it provides them with a more interactive and dynamic programming experience. Instead of following a linear and predetermined development process, they can embrace the flexibility and agility of JITP to rapidly prototype, test, and refine their algorithms during task execution. This real-time feedback loop allows programmers to fine-tune their code based on immediate results and user requirements, leading to more efficient and effective solutions. Additionally, experienced programmers can leverage JITP to explore innovative approaches, as they have the capability to envision and implement complex algorithms on the fly.

Regardless of the user's programming expertise, JITP empowers individuals to be actively involved in the development process, aligning it with their specific goals and requirements. By placing the user at the forefront, JITP fosters a more inclusive computing environment, bridging the gap between users and developers. It encourages users to embrace their algorithmic insights, regardless of their programming background, and provides them with the tools and capabilities to transform these insights into functional and practical software solutions.

## II. JUST-IN-TIME PROGRAMMING PARADIGM

Computing paradigms, such as procedural, object-oriented and functional programming, represent different approaches or models for solving computational problems [3]. These paradigms provide a conceptual framework and guidelines for structuring and organizing the development of software systems. They define the fundamental principles, methodologies, and patterns that shape how computations are performed and how problems are solved within a specific paradigm. Developers often choose a computing framework and paradigm based on the specific requirements of their software projects, the nature of the problem being solved, and the desired trade-offs in terms of performance, scalability, maintainability, and ease of development.

Just-In-Time Programming takes a task-oriented focus, where users concentrate on tasks (and their subtasks), with task completion as the primary goal [2]. During task execution, users envision algorithms to complete subtasks and improve efficiency. JITP enables the immediate implementation of these algorithms, leveraging the user's insights and enhancing task completion in real-time.

JITP offers the following benefits as a computing paradigm:

*1) Dynamic and Adaptive Computing:* JITP addresses the limitations of pre-designed software by enabling users to develop and program tasks while they are in progress. This dynamic and adaptive nature allows for real-time adjustments to meet evolving requirements, making computing more responsive and aligned with immediate user needs. JITP leverages user insights and domain expertise, resulting in tailored solutions that optimize task completion efficiency.

*2) User Empowerment:* JITP places the user at the center of the programming process. Whether the user is a novice or an experienced programmer, JITP enables individuals to recognize algorithmic opportunities during task execution and implement them just in time. This user-centric approach empowers non-programmers to augment software engineers and reduces the reliance on dedicated software development teams, fostering a more inclusive and efficient computing environment.

*3) Increased Productivity:* By developing, implementing and automating potential computer subtasks during task execution, JITP significantly enhances productivity. JITP allows users to capitalize on their algorithmic insights immediately, resulting in faster and more efficient task completion.

*4) Improved Task Understanding and Innovation:* JITP encourages users to gain a deeper understanding of their tasks and subtasks. By actively engaging with the programming process during task execution, users become more aware of the underlying algorithms and automation possibilities within their domain. This heightened understanding can lead to innovative solutions, as users are more likely to identify new approaches and optimize existing ones based on their firsthand experience.

*5) Flexibility and Adaptability:* JITP thrives in environments characterized by rapidly changing requirements and dynamic task execution. As tasks evolve or new insights emerge, users can quickly modify and extend their implemented algorithms to accommodate these changes, fostering a flexible and agile computing framework.

*6) Rapid Prototyping and Iterative Development:* JITP supports rapid prototyping and iterative development. Users can experiment with different algorithms and automation strategies on the fly, testing their effectiveness and refining them iteratively. This iterative development process allows for continuous improvement, reducing the time between idea conception and deployment. JITP's inherent modularity facilitates easy integration and replacement of components, enabling efficient prototyping and experimentation.

*7) Enhanced Error Detection and Debugging:* JITP provides the opportunity for immediate error detection and debugging. Since users are actively involved in the programming process, they can quickly identify and address issues as they arise, minimizing the impact on task completion.

Just-In-Time Programming (JITP) represents a user-centric computing paradigm, empowering users to implement algorithms and develop tasks during task execution. By embracing JITP, users gain the ability to leverage algorithmic insights in real-time, resulting in increased productivity, flexibility, and innovation.

## III. JUST-IN-TIME PROGRAMMING COMPUTING FRAMEWORK

For JITP to achieve mainstream adoption, it is important to develop a framework that provides a structured approach to building software applications and can accommodate any required computation, as it must be responsive to any user input. Here, we demonstrate that integration of flow-based programming techniques and large language models provide an ideal JITP computing framework.

Flow-Based Programming (FBP) offers a structured, modular and reactive workflow model that aligns well with the dynamic nature of task execution and algorithm implementation [4,5]. Similarly, Large Language Models (LLMs) allows for the expressive capacity to represent and manipulate any computable function [6]. The combination of FBP and LLMs allows us to define a general-purpose, Turing-complete programming environment.

### A. Principles of Flow-Based Programming

Flow-Based Programming is a programming paradigm that focuses on the flow of data between components, emphasizing modularity, reusability, and reactive processing. In FBP, the execution of a program is driven by the flow of data, rather than being strictly controlled by a predefined sequence of operations.

The key principles of FBP include:

*1) Component-Based Design:* FBP encourages breaking down a system into smaller, self-contained components. These components have well-defined inputs and outputs, facilitating modularity, code reuse, and easy maintenance.

*2) Data Streams and Connections:* FBP emphasizes the flow of data streams between components. Components can receive input data, process it, and produce output data that is then passed to downstream components. The connections between components define the flow of data, allowing for flexible and reactive execution.

*3) Asynchronous and Reactive Execution:* FBP promotes an asynchronous and reactive execution model. Components react to incoming data, processing it as soon as it becomes available, enabling real-time responsiveness and dynamic task adaptation.

The integration of Flow-Based Programming (FBP) within the JITP framework offers several benefits that enhance task-time development:

*1) Modularity and Reusability:* FBP's component-based design fosters modularity and code reusability. Components can be easily connected and combined, allowing users to create flexible and scalable solutions. This modularity also enables incremental development and iterative improvements, aligning well with the JITP approach.

*2) Dynamic Task Adaptation:* FBP's reactive execution model enables components to react to incoming data in real-time. This flexibility allows for dynamic task adaptation, where the solution can adjust and respond to changing task requirements or data inputs. JITP leverages this adaptability to accommodate evolving user needs and algorithmic insights during task execution.

*3) Scalability and Parallelism:* FBP inherently supports parallel processing and scalability. By leveraging the flow of data between components, tasks can be distributed across multiple processing units, improving performance and efficiency. This scalability is particularly beneficial when dealing with computationally intensive tasks or large datasets.

*4) Visualization and Debugging:* FBP frameworks often provide visual representations of the data flow and component connections, facilitating visualization and debugging of the automation solution. This visual feedback enhances user understanding and aids in identifying and resolving issues during algorithm implementation.

*B. Large Language Models*

Large Language Models (LLMs) are advanced artificial intelligence (AI) models designed to understand and generate human language [7,8,9]. These models, built using deep learning techniques, have been trained on vast amounts of text data from diverse sources such as books, articles, and websites. LLMs excel at tasks such as natural language understanding, text generation, translation, summarization, and have found applications in a wide range of domains, including chatbots, virtual assistants, content generation, language translation, content filtering, and more [10].

LLMs can also be extensively trained on diverse code repositories and documentation, so that the models acquire an understanding of programming syntax, structures, and patterns [11]. LLMs can therefore generate software code by leveraging their language processing capabilities and knowledge of programming concepts. When tasked with generating software code, LLMs can take high-level instructions or prompts provided by users and generate corresponding code snippets or even complete programs. They can analyze the context, infer the desired functionality, and generate code that aligns with the specified requirements.

LLMs can be considered Turing complete [12]. While LLMs are not specifically designed for general-purpose computation like traditional programming languages, they possess the underlying capability to simulate other Turing-complete systems given enough time and resources. This property stems from the expressive capacity of LLMs to represent and manipulate information in the form of text sequences. In the context of JITP, LLMs provide the required "back-end" for a Turing complete JITP environment by responding to user requests, generating code in real time, and seamlessly integrating the code within the rest of the (flow-based) execution.

Integrating Large Language Models (LLMs) with Flow-Based Programming (FBP) can create a powerful framework for Just-In-Time Programming (JITP), combining the capabilities of advanced language models with the modular and reactive workflow of FBP. LLMs leverage their language understanding and code generation capabilities to enable users to express their algorithmic insights and automate tasks in real time. Flow-based programming, with its visual representation of tasks and data flow, provides the overall structured approach by facilitating the incorporation of dynamically generated code into the overall execution workflow.

*C. Framework Implementation Strategy*

To develop an effective JITP framework, we integrate LLMs with FBP in the following way, taking into account certain considerations.

*1) Identify Task-Specific LLMs:* Begin by identifying the LLMs that are most relevant to the specific task domain. Select LLMs that align with the programming language or task requirements to enhance the JITP capabilities.

*2) Define LLM Components:* Next, define LLM components within the FBP framework and design the components to encapsulate the complexity of interacting with the LLMs and provide a simple interface for other components to utilize. These components encapsulate the interactions with LLMs, such as sending input text, retrieving generated code or responses, and managing the LLM state.

*3) Establish Data Flow:* Design the data flow between the LLM components and other components within the FBP framework. Determine the input data required by the LLM component, such as task descriptions, code snippets, or user instructions. Define the outputs from the LLM components, such as generated code, text responses, or relevant suggestions.

*4) Enable Reactive Execution:* Leverage the reactive execution model of FBP to trigger LLM interactions based on incoming data or events. For example, when a user provides a task description or requests assistance, the relevant LLM component can be triggered to generate code that is then subsequently executed. Ensure that the LLM components are reactive and responsive to changes in data inputs, enabling dynamic JITP capabilities.

*5) Handle LLM State Management:* LLMs often have a limited context window, meaning they may not have full access to the entire task history. To overcome this limitation, consider incorporating mechanisms to manage the state of the LLMs. This can involve maintaining a context buffer or session management to provide relevant contextual information to the LLM component during task execution.

*6) Visualize and Debug LLM Interactions:* Utilize visualization and debugging tools provided by the FBP framework to monitor the interactions with LLM components. This enables users to understand the flow of data, identify potential bottlenecks, and troubleshoot any issues related to LLM interactions. Visualization tools can also aid in interpreting LLM-generated outputs and ensuring they align with the desired outcomes.

*7) Iterate and Improve:* Continuously iterate on the LLM integration within the JITP framework based on user feedback, task requirements, and performance evaluation. Refine the LLM components, data flow, and reactive execution to optimize the JITP experience. Incorporate user preferences and algorithmic insights gained during task execution to further enhance the efficiency and effectiveness of the JITP framework.

By integrating LLMs with FBP, developers can leverage the language modeling capabilities of LLMs within the JITP framework, enabling users to generate code, receive suggestions, or obtain relevant information in real-time. This integration combines the strengths of advanced language models with the modularity, scalability, and adaptability of FBP, resulting in a powerful JITP framework capable of supporting a wide range of use cases.

## IV. Composable Just-In-Time Programming Platform

Composable DataOps Platform [1] is a cutting-edge platform initially developed at MIT that is able to integrate Flow-Based Programming and Large Language Models into a robust JITP computing framework. This innovative combination empowers users to leverage the power of LLMs and harness the flexibility of flow-based programming to create, automate, and optimize software solutions in real time.

As a computing framework, Composable provides a structured and comprehensive platform with a set of pre-defined components and built-in tools and libraries for building software applications. By incorporating FBP principles, Composable offers a visual and intuitive environment for designing and orchestrating complex workflows. Users can graphically represent their tasks, data flow, and interactions, enabling a holistic view of their processes. This visual representation simplifies the understanding and manipulation of data pipelines, making it easier to identify bottlenecks, optimize performance, and ensure maintainability.

Flow-based programs in Composable, called "DataFlows", are represented as event-driven workflows, as shown in Figure 1. Each DataFlow consists of one or more Modules that are connected together to produce higher-level functionality. Composable Modules are atomic processing elements with strongly typed inputs and outputs. All information required for a Module to execute is retrieved from its inputs through connections. Modules can be reused easily and interchanged with other Modules. As shown in Figure 1, a Module takes in one or more inputs, and produces one or more outputs. These outputs can then be connected to any number of other Module inputs.

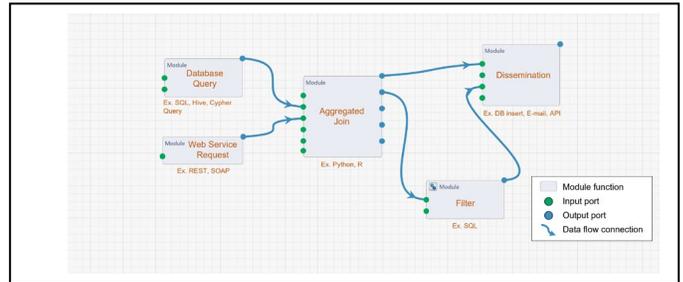

Fig. 1. Schematic representation of a flow-based program, called a DataFlow within the Composable platform. Composable Modules are functional blocks that take in one or many inputs, and produces one or many outputs. These outputs can then be connected to any number of other Module inputs to form a DataFlow.

End-users can compose unique DataFlow Applications by dragging and dropping Modules and connecting them together in an "infinitely configurable" modular design. Composable is designed to handle all data types, and is flexible regarding the consumption of all data sources.

Modules that execute LLM computations can be created. In this article, we demonstrate the use of OpenAI's publicly-available ChatGPT application as the back-end LLM for the JITP framework [13,14]. While ChatGPT is primarily trained on vast amounts of text data and excels at natural language understanding and generation, and other LLMs may be better suited for software code generation, ChatGPT, with its simple API interface [15], is sufficient for our demonstration purposes. Composable is not limited to using just this LLM, and can, in fact, be extended to utilize other purpose-built LLMs [16].

The following sections demonstrate several examples of the Composable JITP implementation.

### A. JIT Code Generation Module

First, we will define a "JIT Code Generation" DataFlow that will serve as an "App Reference" Module (a Module that calls another DataFlow Application) within other DataFlows. The "JIT Code Generation" DataFlow, shown in Figure 2, simply consists of a **WebClient Robust** Module that accepts a single string input as a prompt, makes a http request against the ChatGPT API, and returns the response. Note that while we are using OpenAI's ChatGPT API as our LLM, we can simply "plug in" other LLM implementations.

The **WebClient Robust** Module uses the following parameters:

- Uri: https://api.openai.com/v1/chat/completions

- Method: POST
- Content-Type: application/json
- Header: A **Key Value Pair** Module with Key "Authorization" and Value "Bearer < API _key>"

We externalize one input and two outputs:

- Input: We use an **External String Input** Module, so that we can externalize the input to other DataFlows. The string input is used as a parameter in the **String Formatter** Module, to format the end-user prompt with the syntactically correct json request payload.
- Output: Status code of the web request (e.g., 200).
- Output: String output, after first extracting the json value using the **JSONPath Query** Module, followed by a **Regex Replace** Module. We use the **Regex Replace** Module because ChatGPT usually returns code proceeded with backticks (`), that we use to parse out the actual raw code from the extraneous natural language within the response.

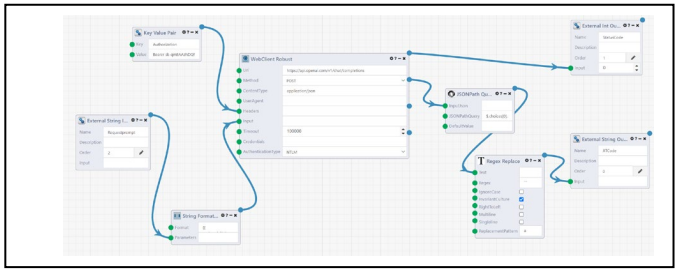

Fig. 2. "JIT Code Generation" DataFlow.

We can use the "JIT Code Generation" DataFlow as an "App Reference" Module within our main execution DataFlow. One of the powerful features of Composable's FBP framework is that DataFlows can be called in this way from within other DataFlows. Figure 3 shows how we can find the newly created DataFlow shown in Figure 2 in the Module Palette, and simply drag and drop it onto the Designer canvas. The App Reference Module shows the single externalized input for the request prompt and the two externalized outputs for the web request status code and raw code text response.

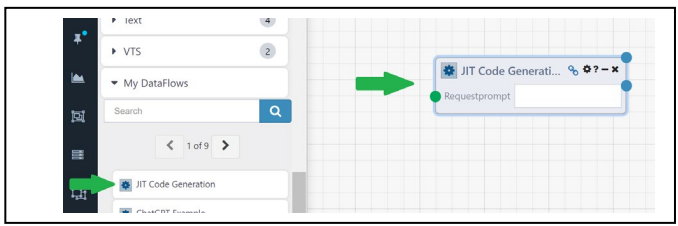

Fig. 3. The "JIT Code Generation" DataFlow can be used as an App Reference Module within a new DataFlow.

### B. Simple, Just-In-Time Arithmetic

As an initial simple Just-In-Time Programming example, we can use the following prompt:

*Write a python function called gptFunction that adds two integers. Only return the raw python code.*

We can add a **Python Code** Module to the DataFlow, which executes a given Python code block with a defined function, as well as two integer inputs, as shown in Figure 4. Figure 5 shows the new output given a slightly altered prompt requesting subtraction rather than addition, showing how easy it is to program just-in-time.

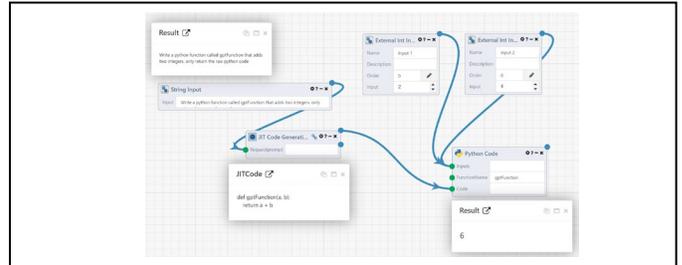

Fig. 4. Simple Just-In-Time Programming DataFlow that requests the addition of two integers, generates code and executes the code for two given inputs. The results after execution is shown for two given inputs.

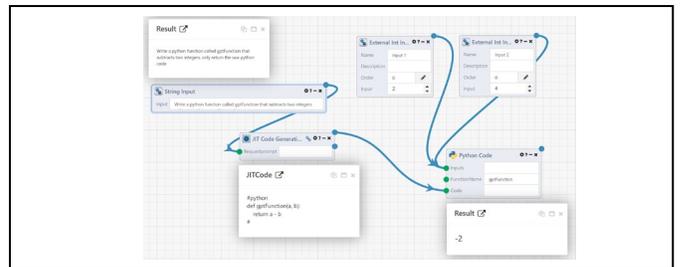

Fig. 5. Simple Just-In-Time Programming DataFlow that requests the subtraction of two integers, generates code and executes the code for two given inputs. The results after execution is shown for two given inputs.

### C. Primality Test

In this next example, we request a just-in-time algorithm for determining whether an input number is prime.

Figure 6 shows a DataFlow that provides a single integer input and uses the following prompt:

*Write a python script that checks if a given command line integer input is prime. Only return the raw python code.*

The "JIT Code Generation" DataFlow returns the Python script shown in Figure 7, with the result, for input 31, given as "31 is prime!".

This DataFlow may be better integrated with other DataFlows if the Primality Test result is simply a Boolean (0 or 1). We can therefore simply adjust the input prompt:

*Write a python script that returns a 1 if a given command line integer input is prime and a 0 if not. Only return the raw python code.*

The new just-in-time code is shown in Figure 8, with the full DataFlow and execution results shown in Figure 9.

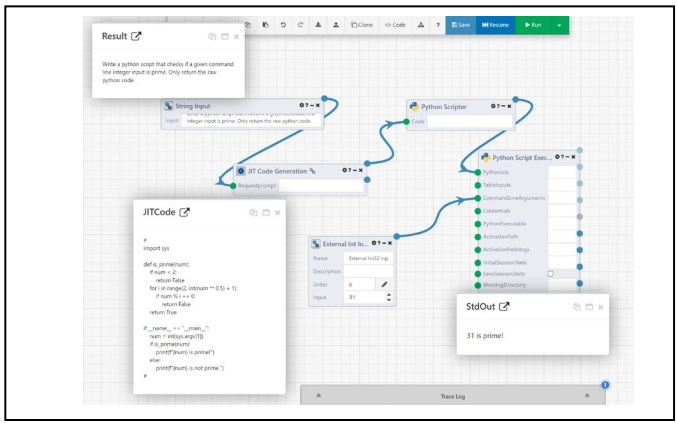

Fig. 6. Primality Test DataFlow showing the just-in-time code being generated and executed.

```
#
import sys

def is_prime(num):
    if num < 2:
        return False
    for i in range(2, int(num ** 0.5) + 1):
        if num % i == 0:
            return False
    return True

if __name__ == "__main__":
    num = int(sys.argv[1])
    if is_prime(num):
        print(f"{num} is prime!")
    else:
        print(f"{num} is not prime.")
#
```

Fig. 7. Just-in-time code snippet that was generated by the LLM for the given input.

```
#
import sys

def is_prime(n):
    if n < 2:
        return 0
    for i in range(2, int(n**0.5)+1):
        if n % i == 0:
            return 0
    return 1

if __name__ == "__main__":
    n = int(sys.argv[1])
    print(is_prime(n))
#
```

Fig. 8. Just-in-time code snippet that was generated by the LLM for the revised input.

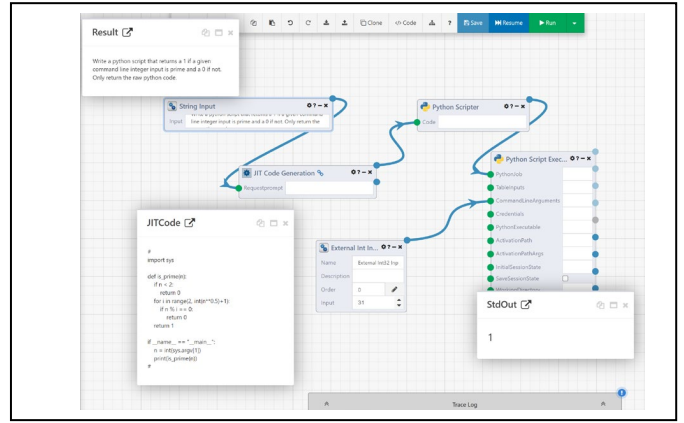

Fig. 9. Revised Primality Test DataFlow showing the just-in-time code being generated and executed.

### D. Generating New Data

In this next example, we show that Just-In-Time Programming can allow a user to generate a dataset.

Here, we use as our prompt:

*Define a pandas dataframe called composable_table_out with column State that contains all States in the USA that border the ocean.*

The "JIT Code Generation" DataFlow returns the Python script shown in Figure 10.

```
#
import pandas as pd

# create a dictionary of States and their ocean borders
states_dict = {'Maine': 'Atlantic', 'New Hampshire': 'Atlantic', 'Massachusetts': 'Atlantic', 'Rhode Island': 'Atlantic', 'Connecticut': 'Atlantic', 'New York': 'Atlantic', 'New Jersey': 'Atlantic', 'Delaware': 'Atlantic', 'Maryland': 'Atlantic', 'Virginia': 'Atlantic', 'North Carolina': 'Atlantic', 'South Carolina': 'Atlantic', 'Georgia': 'Atlantic', 'Florida': 'Atlantic', 'Texas': 'Gulf of Mexico', 'Louisiana': 'Gulf of Mexico', 'Mississippi': 'Gulf of Mexico', 'Alabama': 'Gulf of Mexico', 'California': 'Pacific', 'Oregon': 'Pacific', 'Washington': 'Pacific', 'Alaska': 'Pacific'}

# create a pandas dataframe from the dictionary
composable_table_out = pd.DataFrame(list(states_dict.items()), columns=['State', 'Ocean Border'])

# filter the dataframe to only include States that border the ocean
composable_table_out = composable_table_out[composable_table_out['Ocean Border'].notnull()]

# display the dataframe
print(composable_table_out)
#
```

Fig. 10. Just-in-time code snippet that was generated by the LLM for the input requesting a list of U.S. states that border the ocean.

The DataFlow and table output is shown in Figure 11.

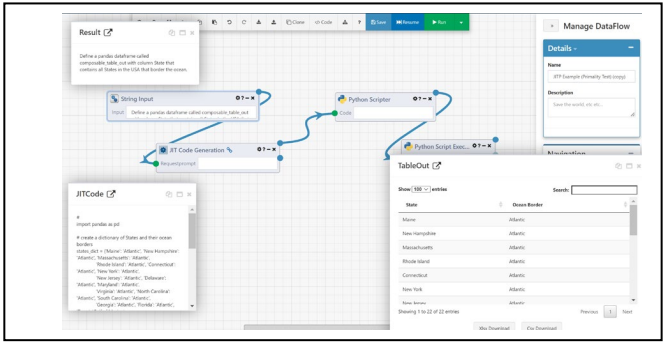

Fig. 11. Generating a new data table using Just-In-Time Programming.

*E. Table Manipulation*

More complex data processing workflows generally require the manipulation of tables. Here we show a request to select only records that appear more than once in an input table.

Our prompt is:

*For a given pandas dataframe called input_dfs[0], define composable_table_out to contain only those records that are duplicates.*

The generated Python is shown in Figure 12, and the DataFlow and results are shown in Figure 13.

```
#python
import pandas as pd

# Assuming input_dfs[0] is your pandas dataframe

# Find duplicate records
duplicates = input_dfs[0][input_dfs[0].duplicated()]

# Create composable_table_out with only duplicate records
composable_table_out = duplicates.copy()

# Display composable_table_out
print(composable_table_out)
#
```

Fig. 12. Just-in-time code snippet that was generated by the LLM for the input requesting the selection of duplicate records in a dataframe.

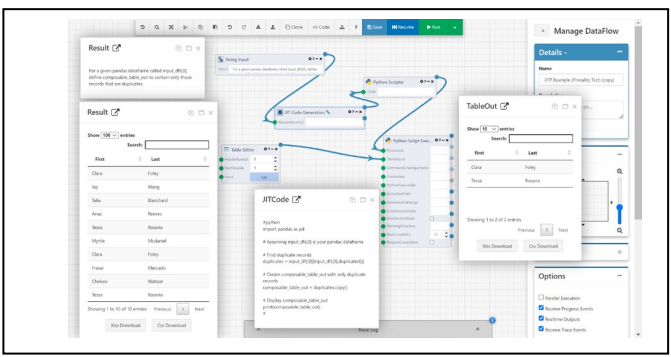

Fig. 13. Table and Data manipulation using JIT programming. We pass in a table with duplicate records, and retrieve only those duplicate records.

## V. Ensuring Trust

While pre-trained foundational LLMs can generate software code, as shown in the above examples with OpenAI's ChatGPT, a language model specifically trained on software source code can provide greater accuracy and contextual understanding. Specifically, models that are trained on a massive dataset of source code can capture code structures and coding conventions and can produce code that is more contextually appropriate, adheres to coding best practices, and aligns with the desired functionality. The domain-specific understanding of a LLM for code can ensure that the generated code is of higher quality and meets the specific requirements of software development tasks, and more practically, the generated responses can contain just raw code, and not any extraneous text or language.

When developing software, even in a just-in-time programming paradigm, we want to ensure trust in the software [17]. In this context, we define trust to mean that the code performs its intended functions correctly and predictably, and that the resulting end-to-end system delivers accurate results, responds to inputs appropriately, and operates without unexpected failures or errors. When the LLM generates just a raw block of text as code to be executed, it is difficult to ensure trust as it requires an expert programmer to read each block of generated code, check it for accuracy, and test it. We see this in the above examples, where the complexity increases from the simple algorithms (for arithmetic operations) to more complex algorithms (primality test) and even more complex data manipulation (finding duplicates).

To overcome this limitation and ensure trust in the JITP framework, we use a LLM to generate not just a block of text to be used as executable code, but rather generate a complete, visual, flow-based program. Specifically, we generate a DataFlow, a visual algorithm composed of pre-defined functional blocks (Modules), to ensure consistency, accuracy and reliability. Our approach leverages two key features of Composable and FBP:

- Strongly Typed Modules: The Composable FBP framework enforces strong typing of Modules, ensuring that data types are explicitly defined and consistent throughout the DataFlow.
- Loose Coupling: The Composable FBP framework promotes loose coupling between Modules, meaning that Modules are decoupled from each other and communicate through well-defined data interfaces.

By generating flow-based programs just in time, we benefit from the following features:

- Visualization of Control Flow: Visual flow-based programs provide a graphical representation of the program's control flow. The flowchart-like diagrams make it easier to comprehend the program's logic and control structures. This visualization aids in understanding the program's intended behavior, making it less prone to errors and enabling better accuracy during development and debugging.
- Clear Representation of Data Flow: Visual flow-based programming emphasizes the flow of data between

different components. By explicitly representing data connections and transformations, it becomes easier to track and validate the data flow within the program. This clarity helps in ensuring correctness and identifying potential issues or bugs related to data handling.

- Reduced Functional Errors: Since the logic is constructed using pre-built functional components, flow-based programs can largely eliminate the potential for underlying errors in the required functions and improve the overall correctness.

- Reduced Cognitive Load: Visual flow-based programming reduces the cognitive load on developers by providing a more intuitive and visual representation of the program. The use of visual blocks and diagrams makes it easier to understand complex logic and relationships between program components. This decreased cognitive load leads to fewer mistakes and better accuracy during development.

- Simpler Debugging Process: When debugging visual flow-based programs, it is often easier to identify and isolate errors. The graphical representation allows developers to visually trace the execution path, track data flow, and identify problematic areas. This ease of debugging helps in identifying and rectifying issues more efficiently, resulting in improved correctness and accuracy.

As an example, we begin with a simple DataFlow that takes two integer inputs, performs an arithmetic computation (addition, subtraction, …), and returns an integer. The DataFlow is shown in Figure 14, and consists of the following Modules:

- Two **External Int Input** Modules for the integer inputs
- A **Calculator Module** to perform the arithmetic computation (e.g., addition)
- An **External Int Output** Module for the integer output

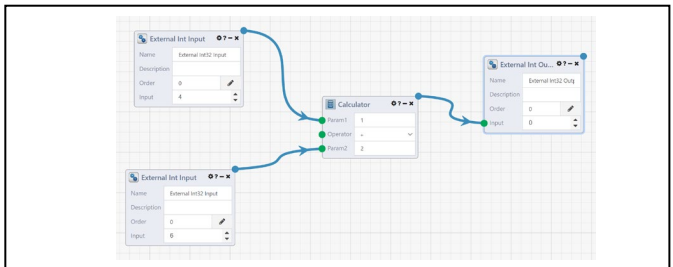

Fig. 14. A simple DataFlow that takes two integers, adds them, and returns an integer.

We are able to convert this visual flow-based program into structured source code (e.g., in C#) using Composable's FluentAPI, a set of C# classes that interact with Composable services [1]. In this way, the DataFlow shown in Figure 14 can be represented as code shown in Figure 15, and the LLM can be trained to generate new DataFlows.

```
//---------------------------------------------------------------
// Fluent Flow Code for the Composable DataOps Platform.
// Database Version: 1.0.339.0
// Assembly Version: 2.0.20885.0
// Composable Build Date: May 11, 2023 10:55:31 AM
// Code Generated Date: June 27, 2023 10:49:25 PM
//---------------------------------------------------------------

using CompAnalytics.Contracts;
using CompAnalytics.FluentAPI;
using System;

public class Program
{
    private static CompAnalytics.IServices.Deploy.ResourceManager CreateManager()
    {
        CompAnalytics.IServices.Deploy.ConnectionSettings connectionSettings =
            new CompAnalytics.IServices.Deploy.ConnectionSettings();
        connectionSettings.Uri = new System.Uri("https://cloud.composableanalytics.com/");
        connectionSettings.AuthMode =
```

Fig. 15. Snippet of code equivalent of the DataFlow shown in Figure 14.

We can see this in action by, for example, enhancing the DataFlow in Figure 14 with the following prompt:

*Generate a program that takes 2 External Integer Inputs, feeds them into a Calculator Module for addition, then feeds it into another Calculator Module along with a third External Integer Inputs for addition, and returns an External Integer Output.*

The generated output is shown in Figure 16 as code and Figure 17 as a visual DataFlow.

```
//---------------------------------------------------------------
using CompAnalytics.Contracts;
using CompAnalytics.FluentAPI;
using System;

public class Program
{
    private static CompAnalytics.IServices.Deploy.ResourceManager CreateManager()
    {
        CompAnalytics.IServices.Deploy.ConnectionSettings connectionSettings = new
```

Fig. 16. Snippet of code generated by the LLM to represent the new DataFlow.

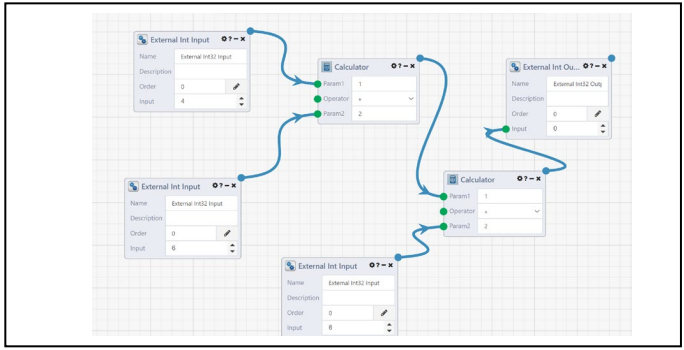

Fig. 17. DataFlow generated by a purpose-built LLM based on a given prompt.

## VI. Conclusion

Just-In-Time Programming offers a user-centric approach to programming by allowing algorithm implementation during task execution. By aligning software functionality with dynamic user requirements, JITP empowers users to leverage their algorithmic insights and implement tasks and subtasks in real-time. The integration of flow-based programming techniques and Large Language Models presents a promising avenue for a JITP framework. Task-oriented focus lies at the core of JITP, with LLMs generating the immediate implementation of algorithms and FBP orchestrating task completion in real-time.

In this paper, we demonstrate the use of Composable as a state-of-the-art JITP computing framework that seamlessly integrates Flow-Based Programming techniques with Large Language Models. By harnessing the power of LLMs within a visual and intuitive environment, Composable enables users to express their algorithmic insights in real time, automate tasks, and rapidly prototype software solutions. The versatility of the Composable JITP Platform extends across various domains and use cases. In data science and analytics, users can leverage LLMs to generate code for data preprocessing, feature engineering, and model evaluation, while orchestrating complex data workflows with FBP principles. In software development, Composable facilitates rapid prototyping, automates repetitive tasks, and allows for the development of large, microservices-based architectures, with seamless integration of LLM-generated code within the larger codebase. Additionally, Composable finds applications in natural language processing, machine learning, robotic process automation, and more, where the combination of LLMs and FBP principles offers unparalleled flexibility and agility.